\documentclass[10pt,letterpaper,twocolumn]{article} 

\usepackage{ol}
\usepackage{hyperref}
\usepackage{amsmath}

\begin{document}

\twocolumn[ 
\title{Channel plasmon-polaritons: modal shape, dispersion, and losses}
\author{Esteban Moreno and F. J. Garcia-Vidal}
\address{Departamento de F\'isica Te\'orica de la Materia
Condensada, Universidad Aut\'onoma de Madrid, E-28049 Madrid,
Spain}

\author{Sergio G. Rodrigo and L. Martin-Moreno}
\address{Departamento de F\'isica de la Materia Condensada,
Universidad de Zaragoza-CSIC, E-50009 Zaragoza, Spain}

\author{Sergey I. Bozhevolnyi}
\address{Department of Physics and Nanotechnology,
Aalborg University, DK-9220 Aalborg Ost, Denmark}

\begin{abstract}
We theoretically study channel plasmon-polaritons (CPPs) with a
geometry similar to that in recent experiments at telecom
wavelengths (Bozhevolnyi \textit{et al}., Nature {\bf 440}, 508
(2006)). The CPP modal shape, dispersion relation, and losses are
simulated using the multiple multipole method and the finite
difference time domain technique. It is shown that, with the
increase of the wavelength, the fundamental CPP mode shifts
progressively towards the groove opening, ceasing to be guided at
the groove bottom and becoming hybridized with wedge
plasmon-polaritons running along the groove edges.\end{abstract}

] 

The guiding of light within a subwavelength cross section has been
recently attracting a great deal of attention due to ever
increasing demands for miniaturization of photonic circuits. Light
may be confined in the direction perpendicular to a flat metallic
surface for energies below the metal plasma frequency. The mode
guided along the metallic interface is known as surface
plasmon-polariton~(SPP). Various geometries have been proposed to
achieve confinement of the plasmon-polariton in the plane
transverse to the propagation
direction\cite{takahara97,berini99,pile04,pile05a,pile05b}. Among
these proposals, the plasmon-polariton guided by a V-shaped groove
carved in a metal (channel plasmon-polariton, CPP) is particularly
interesting. CPPs were theoretically suggested by Maradudin and
coworkers\cite{maradudin02} and subsequently studied in the
visible regime\cite{pile04,pile05c}. Recently, CPPs have been
experimentally investigated at telecom
wavelengths\cite{bozhevolnyi05}, displaying strong confinement,
low damping, and robustness against channel bending. Thank to
these properties, prototypes of basic devices could be
demonstrated\cite{bozhevolnyi06}. The mentioned devices have been
developed with the help of the effective index approximation but,
to our knowledge, no rigorous electrodynamic computation of CPPs
at telecom wavelengths has been reported. The effective index
approximation can deliver information about the dispersion
relation, but it is expected to be inaccurate for frequencies
close to the mode cutoff and is unable to determine modal shape
and polarization. The functionality of many devices relies on the
overlapping of electromagnetic fields at various sites inside the
device. For this reason the knowledge of the modal shape is
essential to provide a solid foundation for the design of
CPP-based devices. Here we present rigorous simulations of guided
CPPs aimed to elucidate their characteristics at telecom
wavelengths, including full vectorial modes, dispersion, and
losses. We show that, contrary to what is commonly believed, CPPs
at telecom wavelengths are not guided at the groove bottom, at
least for the groove parameters used in the
experiments\cite{bozhevolnyi05,bozhevolnyi06}. Instead, the CPP
field at the groove entrance hybridizes with wedge
plasmon-polaritons (WPPs) running along the edges of the groove.

Our goal is to understand the fundamental CPP mode guided by
realistic grooves at telecom wavelengths\cite{bozhevolnyi05}.
Nevertheless, in order to comprehend the behavior in this regime,
which is close to cutoff, we will consider a broader spectrum,
higher order modes, and a number of different geometries. The
simulations have been performed with two rigorous electrodynamic
techniques: the multiple multipole method (MMP)\cite{hafner99}
and, where mentioned, the finite difference time domain (FDTD)
method\cite{taflove00}. Within the MMP method the corners are
rounded ($10\, \textrm{nm}$ radius of curvature). FDTD results
were converged for a mesh of about $5\, \textrm{nm}$. Such fine
meshes are essential, the more so for wavelengths shorter than
$0.8\, \mu\textrm{m}$. The grooves are carved in gold and we
employ experimentally measured values of the dielectric
permittivity $\varepsilon$.

\begin{figure}[t]
\centerline{\includegraphics[width=8.4cm]{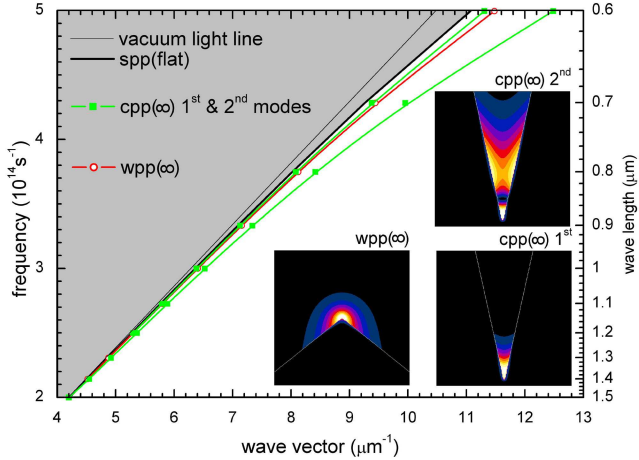}}
\caption{(Color online) Dispersion relation for various modes.
Black thick line: SPP mode on a flat surface. Green lines
(squares): CPP($\infty$) modes for an infinitely deep groove. Red
line (open circles): WPP($\infty$) mode for an infinitely deep
wedge. Right insets: time averaged electric field of the two
CPP($\infty$) modes at $0.6\, \mu\textrm{m}$. Left inset: same for
the WPP($\infty$) mode. The lateral size of the insets is $2\,
\mu\textrm{m}$.}
\end{figure}

Figure~1 shows the dispersion relation for a non-truncated groove
with a semiangle of $\theta=12.5^\circ$ and infinitely long sides.
This structure sustains two modes, being termed CPP($\infty$) (see
right insets), which are outside the dispersion line of the SPP at
a flat surface. The modal shape (time averaged electric field) is
shown in the right insets for a wavelength of $\lambda=0.6\,
\mu\textrm{m}$. In the figure it is also plotted the dispersion
relation for a non-truncated metallic wedge of semiangle
$\alpha=51.25^\circ$ and infinitely long sides. The corresponding
wedge mode running along the edge is termed WPP($\infty$) (see
left inset). WPP($\infty$) for this $\alpha$ will be relevant when
we later truncate the above groove at a finite height: it
corresponds to the edges at both sides of the finite-height
groove. The WPP($\infty$) modal field at $0.6\, \mu\textrm{m}$ is
shown in the left inset. For increasing wavelength all three modes
approach the SPP line (none of them has a cutoff). In this process
the modal shapes remain qualitatively the same, the only
difference being that the fields are expelled away from the groove
or wedge corners.

\begin{figure}[t]
\centerline{\includegraphics[width=8.4cm]{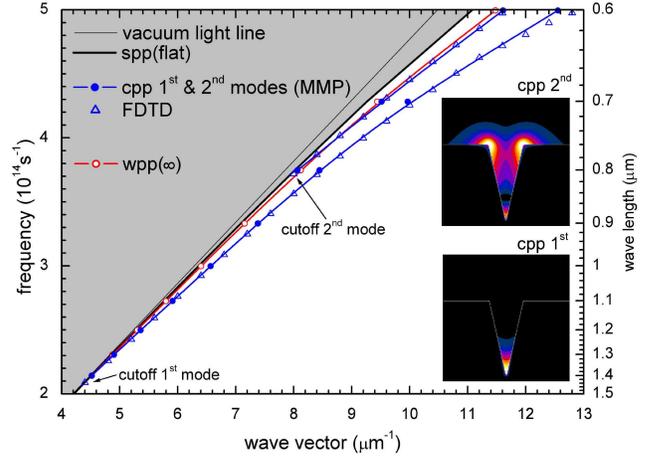}}
\caption{(Color online) Dispersion relation for various modes.
Black thick line: SPP mode on a flat surface. Blue lines (full
circles): CPP modes for a groove of height $1.172\, \mu\textrm{m}$
(computed with MMP method). Triangles: same as before computed
with FDTD method. Red line (open circles): WPP($\infty$) mode for
an infinitely deep wedge. Insets: time averaged electric field of
the two CPP modes at $0.6\, \mu\textrm{m}$. The lateral size of
the insets is $2\, \mu\textrm{m}$.}
\end{figure}

Figure~2 represents a similar plot but now a groove of finite
height is considered, the height being $1.172\, \mu\textrm{m}$.
The CPP modes exhibit now cutoff at different wavelengths ($\sim
1.44\,\mu\textrm{m}$ for the first mode and $\sim
0.82\,\mu\textrm{m}$ for the second one). This was advanced in
Ref.~\onlinecite{gramotnev04}, and it is a consequence of the
above mentioned behavior of the fields for increasing wavelength.
As the wavelength grows, the field is pushed out of the groove
and, after a certain threshold, it can no longer be confined by
the groove sides and is radiated in the form of SPPs along the
contiguous horizontal metal surfaces. It is important to realize
that, before reaching the SPP dispersion line, both modes approach
and cross the WPP($\infty$) line. This means that close to cutoff
the CPP modes must be hybridized with the modes running on the
edges at both sides of the groove. This idea is visualized in the
insets, that render the modal shapes (time averaged electric
field) at $0.6\, \mu\textrm{m}$. At this wavelength the first mode
is not close to WPP($\infty$) and the hybridization does not take
place, but it is already happening for the second mode. The
described phenomenon is even more distinct in Fig.~3 displaying
the fundamental mode for increasing wavelengths. It is observed
that the CPP mode becomes more and more mixed with the
WPP($\infty$). Close to cutoff (at about $1.44\,\mu\textrm{m}$)
the mode is not guided at the groove bottom anymore but rather at
the groove edges. A hint of this possibility was mentioned in
Ref.~\onlinecite{volkov06}. In the experiments, the edges at both
sides of the groove have larger radius of curvature than in the
previously presented simulations. We have verified that this does
not alter our conclusion by repeating the same computation with a
radius of curvature of $100\, \textrm{nm}$ at the groove edges
(and keeping $10\, \textrm{nm}$ at the bottom). Figure~3(d) shows
the instantaneous transverse electric field for this case and it
is clear that hybridization with edge modes still occurs. The
transverse electric field is approximately horizontal inside the
channel (an assumption used by the effective index approximation),
but it is not horizontal near the edges where the field is
maximum. Let us note in passing the excellent agreement of the two
techniques employed here (the residual discrepancy in Fig.~2 for
the fundamental mode at $0.6\, \mu\textrm{m}$ is due to different
rounding schemes of the groove bottom in the two methods). From
the point of view of fabrication it is useful to mention that, for
$\lambda \in (0.6\,\mu\textrm{m},0.8\,\mu\textrm{m})$, the
dispersion relation is extremely sensitive to the fine details of
the groove bottom (e.g., rounding), as concluded after a large
number of simulations where the details of the bottom were
subjected to small perturbations. On the other hand, this does not
happen for telecom wavelengths (as expected from the modal shape),
a circumstance that has also been observed
experimentally\cite{bozhevolnyi06}. Note that the calculated
cutoff wavelength of the fundamental mode is somewhat lower than
the wavelengths used in the experiments. This discrepancy can be
ascribed to (small) differences in the groove geometry, both in
the groove shape (angle, side flatness) and in the groove depth,
and/or different dielectric permittivity of gold.
 We have verified (not shown here for brevity) that
slightly less negative $\varepsilon$ or/and smaller groove
semiangle $\theta$ leads to a higher cutoff wavelength. Finally,
the experiments were conducted at ambient conditions so that water
condensation could not be excluded (a very thin water layer can
significantly increase the cutoff wavelength).

\begin{figure}[t]
\centerline{\includegraphics[width=8.4cm]{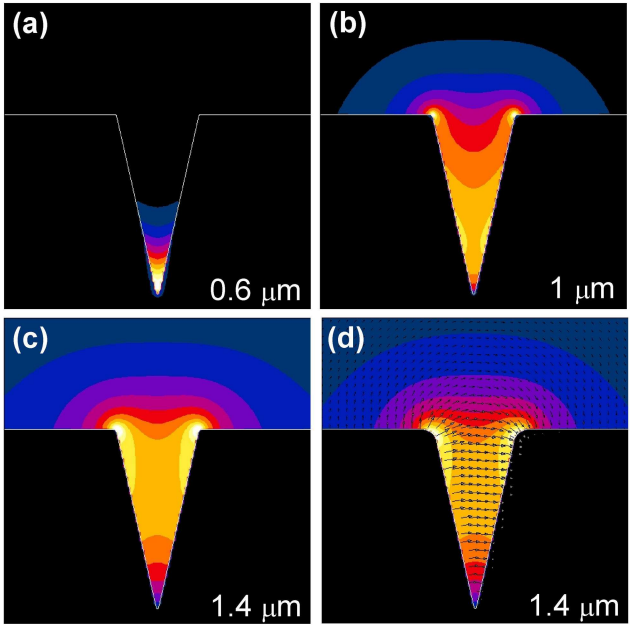}}
\caption{(Color online) Modal shape of the CPP fundamental mode
for increasing wavelength $\lambda$. (a) $\lambda=0.6\,
\mu\textrm{m}$, (b) $\lambda=1\, \mu\textrm{m}$,  (c)
$\lambda=1.4\, \mu\textrm{m}$ (close to cutoff). These panels
display the time averaged electric field. (d) Instantaneous
transverse electric field at $\lambda=1.4\, \mu\textrm{m}$ for a
structure with groove edges rounded with $100\, \textrm{nm}$
radius of curvature. All panels have a lateral size of $2\,
\mu\textrm{m}$.}
\end{figure}

The effect of absorption is summarized in Fig.~4 that renders the
propagation length $l=[2\textrm{Im}(k_z)]^{-1}$, versus wavelength
for the various considered structures ($k_z$ is the modal wave
vector). The propagation lengths are in all cases smaller than
that of SPPs at a flat surface. This is a consequence of the field
enhancement at the corners and the field confinement that
decreases the portion of field propagating in air. When comparing
the CPP modes it is observed that the effect of truncation at a
finite height is only important for wavelengths larger than
$1\,\mu\textrm{m}$, which is reasonable because the field is very
much confined at the groove bottom for smaller $\lambda$. For
longer wavelengths the CPP propagation length is decreased as
compared to that of CPP($\infty$). At $\lambda=1.4\,\mu\textrm{m}$
we find that $l_{\textrm{CPP}}=53\,\mu\textrm{m}$. The values
reported in Ref.~\onlinecite{bozhevolnyi05} at
$1.55\,\mu\textrm{m}$ are twice as large. The discrepancy can be
again ascribed to slight differences in geometry and/or dielectric
permittivity that rise the cutoff wavelength. If the trend of the
line corresponding to the CPP is extrapolated, we find good
agreement with the reported data. It is to be observed that the
propagation length of WPP($\infty$) is significantly higher, a
fact that could find obvious applications.

\begin{figure}[t]
\centerline{\includegraphics[width=8.4cm]{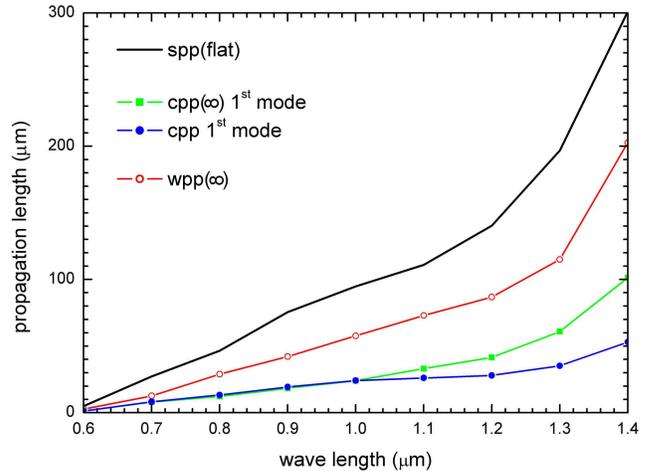}}
\caption{(Color online) Propagation length versus wavelength for
various modes. Black thick line: SPP mode on a flat surface. Blue
line (full circles): CPP fundamental mode for a groove of height
$1.172\, \mu\textrm{m}$. Green line (squares): CPP($\infty$)
fundamental mode for an infinitely deep groove. Red line (open
circles): WPP($\infty$) mode for an infinitely deep wedge.}
\end{figure}

In conclusion, we have presented rigorous computer simulations of
CPPs at telecom wavelength. CPPs have been fully characterized in
terms of modal shape, dispersion, and losses. We have shown that,
for relatively shallow grooves, the field is guided at the groove
opening and is hybridized with modes running along the groove
edges (WPPs). We expect that our findings will be of help for the
design of improved CPP devices.



\begin{thebibliography}{99}

\bibitem{takahara97} J. Takahara, S. Yamagishi, H. Taki, A. Morimoto, and T. Kobayashi, Opt. Lett. {\bf 22}, 475
(1997).

\bibitem{berini99} P. Berini, Opt. Lett {\bf 24}, 1011 (1999).

\bibitem{pile04} D. F. P. Pile and D. K. Gramotnev, Opt. Lett. {\bf 29}, 1069 (2004).

\bibitem{pile05a} D. F. P. Pile, T. Ogawa, D. K. Gramotnev, T. Okamoto, M. Haraguchi, M. Fukui, and S. Matsuo, Appl. Phys. Lett.
{\bf 87}, 061106 (2005).

\bibitem{pile05b} D. F. P. Pile, T. Ogawa, D. K. Gramotnev, Y. Matsuzaki, K. C. Vernon, K. Yamaguchi, T. Okamoto, M. Haraguchi, and M. Fukui, Appl. Phys. Lett.
{\bf 87}, 261114 (2005).

\bibitem{maradudin02} I. V. Novikov and A. A. Maradudin, Phys. Rev. B {\bf 66},
035403 (2002).

\bibitem{pile05c} D. F. P. Pile and D. K. Gramotnev, Opt. Lett. {\bf 30}, 1186 (2005).

\bibitem{bozhevolnyi05} S. I. Bozhevolnyi, V. S. Volkov, E. Devaux, and T. W. Ebbesen, Phys. Rev. Lett {\bf 95}, 046802 (2005).

\bibitem{bozhevolnyi06} S. I. Bozhevolnyi, V. S. Volkov, E. Devaux, J.-Y. Laluet, and T. W. Ebbesen, Nature {\bf 440}, 508 (2006).

\bibitem{hafner99} C. Hafner, {\it Post-Modern Electromagnetics}
(Wiley, Chichester, 1999).

\bibitem{taflove00} A. Taflove and S. Hagness, {\it Computational Electrodynamics: The Finite-Difference Time-Domain Method}
(Artech House, Boston, 2000).

\bibitem{gramotnev04} D. K. Gramotnev and D. F. P. Pile,
Appl. Phys. Lett. {\bf 85}, 266323 (2004).

\bibitem{volkov06} V. S. Volkov, S. I. Bozhevolnyi, E. Devaux, and T. W. Ebbesen, Opt. Express {\bf 14}, 4494 (2006).

\end{thebibliography}
\end{document}